\documentclass[showkeys,preprint,showpacs,preprintnumbers, linenumbers ,amsmath,amssymb]{revtex4-1}
\usepackage{multirow}

\usepackage{graphics}%
\usepackage{epsfig}%
\usepackage{subfigure}

\usepackage[latin1]{inputenc}
\usepackage{t1enc}

\usepackage{algorithmic}

\begin{document}
%\linenumbers

\title{GPUMCD: a new GPU-Oriented Monte Carlo dose calculation platform}% Force line breaks with \\

\author{Sami Hissoiny}
\email{sami.hissoiny@polymtl.ca}
\affiliation{École polytechnique de Montréal, Département de génie
informatique et génie logiciel, 2500 chemin de Polytechnique.
Montréal (Québec), CANADA H3T 1J4}
\author{Hugo Bouchard}
\affiliation{Département de Physique, Université de Montréal, Pavillon Roger-Gaudry (D-428), 2900 
Boulevard Édouard-Montpetit, Montréal, Québec H3T 1J4, Canada and Département de Radio-oncologie, 
Centre hospitalier de l'Université de Montréal (CHUM), 1560 Sherbrooke est, Montréal, Québec H2L 4M1, 
Canada}
\author{Benoît Ozell}
\affiliation{École polytechnique de Montréal, Département de génie
informatique et génie logiciel, 2500 chemin de Polytechnique.
Montréal (Québec), CANADA H3T 1J4}
\author{Philippe Després}
\affiliation{Département de Physique, Université de Montréal, Pavillon Roger-Gaudry (D-428), 2900 
Boulevard Édouard-Montpetit, Montréal, Québec H3T 1J4, Canada and Département de Radio-oncologie, 
Centre hospitalier de l'Université de Montréal (CHUM), 1560 Sherbrooke est, Montréal, Québec H2L 4M1, 
Canada}

\date{\today}% It is always \today, today,

\begin{abstract}\noindent
{\bf Purpose :} Monte Carlo methods are considered the gold standard for dosimetric computations in 
radiotherapy. Their execution time is however still an obstacle to the routine use of Monte Carlo packages 
in a clinical setting.  To address this problem, a completely new, and designed from the ground up for the GPU, Monte Carlo dose calculation package 
for voxelized geometries is proposed: GPUMCD.
\newline {\bf Method :} GPUMCD implements a coupled photon-electron Monte Carlo simulation for 
energies in the range 0.01~MeV to 20~MeV. An analogue simulation of photon interactions is used and a 
Class II condensed history method has been implemented for the simulation of electrons. A new GPU 
random number generator, some divergence reduction methods as well as other optimization strategies are 
also described. GPUMCD was run on a NVIDIA GTX480 while single threaded implementations of EGSnrc and DPM were run on an Intel Core i7 860. 
\newline {\bf Results :} Dosimetric results obtained with GPUMCD were compared to EGSnrc. In all but one 
test case, 98\% or more of all significant voxels passed a gamma criteria of 2\%-2mm. In terms of execution 
speed and efficiency, GPUMCD is more than 900 times faster than EGSnrc and more than 200 times faster than DPM, 
a Monte Carlo package aiming fast executions. Absolute execution times of less than 0.3~s are found for 
the simulation of 1M electrons and 4M photons in water for monoenergetic beams of 15~MeV, including GPU-CPU memory transfers.
%verifier les chiffres
\newline {\bf Conclusion :} GPUMCD, a new GPU-oriented Monte Carlo dose calculation platform, has been 
compared to EGSnrc and DPM in terms of dosimetric results and execution speed. Its accuracy and speed 
make it an interesting solution for full Monte Carlo dose calculation in radiation oncology.

\end{abstract}

%Use showkeys class option if keyword
\maketitle

\section{\label{sec:intro}Introduction}
%% ces packages ne sont pas forcément dédiés au calcul de dose
Monte Carlo packages such as EGS4~\cite{NELS85}, EGSnrc~\cite{KAWR00,KAWR03}, EGS5~
\cite{HIRA05} and PENELOPE~\cite{SALV06,BARO95} have been extensively compared to experimental data in a large array of conditions, and generally demonstrate excellent agreement with measurements. EGSnrc, for instance, was shown to ``pass the Fano cavity test at the 0.1\% level''~\cite{ROGE06}. Despite this accuracy, Monte 
Carlo platforms are mostly absent from routine dose calculation in radiation therapy due to 
the long computation time required to achieve sufficient statistical significance.

Monte Carlo simulations are generally considered the gold standard for benchmarking analytical dose 
calculation approaches such as pencil-beam based computations~\cite{MOHA86,AHNE92,JELE05} and 
convolution-superposition techniques~\cite{MACK85,LIU97}. Such techniques typically use precomputed 
Monte Carlo data to incorporate physics-rich elements in the dose calculation process. These semi-empirical methods were developed as a trade-off between accuracy and computation time and as such do 
not match the level of accuracy offered by Monte Carlo simulations, especially in complex heterogeneous geometries 
where differences of up to 10\% were reported~\cite{KRIE05,DING05,MA00}.

Fast Monte Carlo platforms have been developed for the specific purpose of 
radiotherapy dose calculations, namely VMC~\cite{KAWR96,FIPP99,GARD07} and DPM~\cite{SEMP00}. 
These packages make some sacrifices to the generality and absolute accuracy of the simulation by, for 
example, restricting the energy range of incoming particles. This assumption allows simpler 
treatment of certain particle interactions which are less relevant within the limited energy range considered. Gains in 
efficiency of around 50 times can be achieved by these packages when compared to general-purpose 
solutions.

This paper presents GPUMCD (GPU Monte Carlo Dose), a new code that follows the fast Monte Carlo 
approach. GPUMCD relies on a relatively new hardware platform for computations: the GPU (Graphics 
Processing Unit). The GPU is gaining momentum in medical physics~\cite{KUTT09a, JIA10b, 
GREE09,HISS10} as well as in other spheres~\cite{ANDE08,CHEN09,PREI09,VOLK08,JANU10} 
where high-performance computing is required. A layer-oriented simulation based on the MCML Monte 
Carlo code for photon transport has already been ported to the GPU~\cite{ALER08}. The photon transport 
algorithm from PENELOPE has also been ported to the GPU~\cite{BADA09} for accelerations of up to 27x. 
DPM~\cite{SEMP00} has been ported to the GPU with excellent dosimetric results compared 
to the CPU version, as expected, but with
relatively low (5-6.6x) acceleration factors~\cite{JIA10}.

%% je mettrais ça dans la discussion non?
A comparison of GPUMCD to the work by Jia {\em et al.} must be made cautiously. The work of Jia {\em et al.} is a port of an existing algorithm that is not, to the best of our knowledge, designed for the specific nature of the GPU. Notable differences are found between the two approaches regarding the treatment of secondary particles and the electron-photon coupling. These differences will be exposed in Sec.~\ref{sec:matred}. The code presented in this article is the first attempt of a complete rewrite of a coupled photon-electron 
Monte Carlo code specifically designed for the GPU. New techniques for the memory management of particles as well as efforts to reduce the inherent divergent nature of Monte Carlo algorithms are detailed. Divergence in GPU programming arises from conditional branching, which is not optimal in stream processing where predictable execution is expected for optimal performances. No new sampling methods used in Monte Carlo simulations were introduced in GPUMCD; all theoretical developments of cross section sampling 
methods have been adapted from their description in general-purpose Monte Carlo package manuals (and 
references therein) listed previously. However, their actual implementation is original, as it is specifically tailored 
for parallel execution on graphics hardware, and the code has not been taken from existing Monte Carlo implementations.
%Also, theoretical cross-section equations will later be presented but the actual cross section data used in 
%GPUMCD comes from PEGS4.
%% fin de  je mettrais ça dans la discussion non?

The paper is structured as follows: Sec.~\ref{sec:material} introduces the physics principles as well as the 
hardware platform and implementation details. Sec.~\ref{sec:results} presents the results in terms of dose 
and calculation efficiency of GPUMCD compared to the EGSnrc and DPM platforms. Finally, a discussion of 
the reported results is presented in  Sec.~\ref{sec:discussion}.

\section{\label{sec:material}Material and methods}
%inutile selon moi
%In this chapter, the physics and transport algorithms for photons and electrons are presented in Sec.~
%\ref{sec:matphoton} and~\ref{sec:matelectron}. We also present the overall architecture of the new platform 
%in Sec.~\ref{sec:matarchi} as well ass GPU-centric implementation details in Sec.~\ref{sec:matgpu}. Finally, 
%the performance evaluation methodology that has been used to compare the platforms is presented in 
%Sec.~\ref{sec:matperf}.

\subsection{\label{sec:matphoton}Photons simulation}
Photon transport can be modeled with an analogue simulation,~\emph{i.e.} every interaction is 
modeled independently until the particle leaves the geometry or its energy falls below a certain energy level 
referenced as $P_{cut}$ from now on. This allows for an easy implementation of the transport process.

%% as-tu une référence pour ton safely ignored?
GPUMCD takes into account Compton scattering, the photoelectric effect and pair production. Rayleigh 
scattering can be safely ignored for the energy range considered (10~keV-20~MeV). Considering a 
homogeneous phantom, the distance to the next interaction, noted $s$, can be sampled from the following 
expression:
\begin{equation}\label{eqdistance}
s=-\frac{1}{\mu(E)}\rm{ln}(\zeta),
\end{equation}
where $\zeta$ is a random variable uniformly distributed in (0,1] and $\mu(E)$ is the value of the total 
attenuation coefficient, given by

\begin{align}
\mu= \frac{N_A}{A}\rho\left(\sigma_{compton}+\sigma_{photo}+\sigma_{pair}\right),
\end{align}
where the $\sigma$'s are total cross section values for the corresponding interactions, $N_A$ is Avogadro's 
number,  $A$ is the molecular weight and $\rho$ is the density.

To eliminate the need for a distinct geometry engine in heterogeneous situations, GPUMCD employs the 
Woodcock raytracing algorithm~\cite{CART72} in which the volume can be considered homogeneously 
attenuating by introducing a fictitious attenuation coefficient, corresponding to a fictitious interaction which 
leaves the direction and energy of the particle unchanged, in every voxel $\vec{x}$:
\begin{equation}
\mu(\vec{x})_{fict} = \mu_{max} - \mu(\vec{x})
\end{equation}
where $\mu_{max}$ is the attenuation coefficient of the most attenuating voxel inside the volume. The 
distance to the next interaction is therefore always sampled using $\mu_{max}$. The sampling of the 
interaction type, once at the interaction position, is then performed taking into consideration this fictitious 
interaction type. This method is not an approximation, it leads to the correct result even in arbitrarily heterogeneous geometries.

Compton scattering is modeled with a free electron approximation using the Klein-Nishina cross section~
\cite{KLEI29}. The energy and direction of the scattered photon and secondary electron are sampled 
according to the direct method derived by Everett ~\emph{et al.}~\cite{EVER71}. Binding effects and atomic 
relaxation are not modeled. These approximations are reasonable when the particle energy is large compared to the electron binding energy which is the case for radiation therapy~\cite{JOHN83a}.

The photoelectric effect is modeled by once again ignoring atomic relaxation. Additionally, shell sampling is also ignored and all electrons are assumed to be ejected from the K-shell. The sole product of the 
interaction is a new electron in motion with total energy $E_{elec} = E_{phot}$. The angular sampling of the 
electron is selected according to the Sauter distribution~\cite{SAUT31} following the sampling formula 
presented in Sec. 2.2 of the PENELOPE manual~\cite{SALV06}. The simulation of the photoelectric effect will become less accurate when the energy of the photon is in the order of the characteristic X-ray radiation energy of the material since characteristic X-rays are not produced. Therefore, at the previously stated lower energy range threshold of 10~keV, photons are still tracked and simulated but the accuracy of the simulation of the photoelectric effect will be dependent on the material where the simulation is taking place. The binding energies being low in organic compound, this approximation is considered acceptable.

Relatively crude approximations are employed when simulating pair production events. First and foremost, 
no positrons are simulated in this package. The pair production event instead generates two secondary 
electrons. The relatively low probability of pair production events at the energies considered as well as the additional low probability of in-flight annihilation of the positron make this approximation suitable~\cite{SEMP00}. Triplet production is also not modeled. The energy of the incoming photon is trivially split 
between the two electrons using a uniformly distributed random number. The angle of both electrons is 
sampled using the algorithm presented in Eq. 2.1.18 of the EGSnrc manual~\cite{KAWR06}.

\subsection{\label{sec:matelectron}Electron simulation}
Because of the much larger number of interactions experienced by a charged particle before it has deposited 
all of its energy, analogue simulations cannot be used practically. A class II condensed history method is 
used here, as defined by Berger~\cite{BERG63}, in which interactions are simulated explicitly only if they cause a 
change in orientation greater than $\theta_c$ or a change in energy larger than $E_c$. Below these 
thresholds, the interactions are modeled as taking part of a larger condensed interaction, the result of 
which is a major change in direction and energy. Above the thresholds, interactions, usually called hard or 
catastrophic interactions, are modeled in an analogue manner similar to the simulation of photons.
%prochaine phrase non-essentielle
%Hard interactions will be presented in Sec.~\ref{sec:mathardint}, multiple scattering aspects of the 
%condensed history method in Sec.~\ref{sec:matcond} and a general overview of the electron transport 
%mechanic is given in Sec~\ref{sec:mattrans}.

\subsubsection{\label{sec:mathardint}Hard interactions}
GPUMCD simulates inelastic collisions as well as bremsstrahlung in an analogue manner, as long as the 
changes to the state of the particle are greater than the previously discussed threshold.

For inelastic collisions, a free electron approximation is used and no electron impact ionization nor spin 
effects are modeled. The Møller cross section is used to describe the change in energy and direction of the 
scattered and knock-on electrons. The sampling routine presented in Sec. 2.4.3.i of the EGSnrc manual~
\cite{KAWR06} is used.

During a bremsstrahlung event, the energy of the created photon is sampled as presented in Sec. 2.4.2.ii of 
the EGSnrc manual. Approximations to the angular events are employed : the angle of the electron is 
unchanged and the angle of the photon is selected as $\theta = m_0c^2/E$ where $E$ is the energy of the 
incoming electron.

\subsubsection{\label{sec:matcond}Multiple scattering}

The class II condensed history method requires the selection of an angle for the multiply-scattered electron 
after a step, as discussed in the next subsection. To this end, the formulation presented by Kawrakow and 
Bielajew~\cite{KAWR98} and Kawrakow~\cite{KAWR00} is used. The required $q_{SR}^{2+}$ data are 
imported from the \texttt{msnew.dat} file packaged with EGSnrc.

\subsubsection{\label{sec:mattrans}Electrons transport}
%voir commentaire hugo pour cette phrase
The transport of electrons is less straightforward than photon transport because of the high 
probability of soft collisions, which are regrouped in a single step between two hard 
interactions (collision or bremsstrahlung events). This condensed history method
introduces an unphysical (but mathematically converging) factor for the length of the multiple-scattering 
step. Between consecutive hard collisions, which are separated by a distance sampled with an electron version of 
Eq.~(\ref{eqdistance}), a number of multiple-scattering events will occur. In GPUMCD, a simple step-length 
selection (henceforth $e_s$) is used and can be summarized as
\begin{equation}
e_s = min\{d_{vox},e_{s-max},e_{hard} \}
\end{equation}
where $d_{vox}$ is the distance to the next voxel boundary, $e_{s-max}$ is a user-supplied upper bound 
of step length and $e_{hard}$ is the distance to the next hard interaction.

%With this approach, the electron transport process is presented schematically in Fig~\ref{fig:electran} and 
%will be detailed in the following paragraphs.
%
%\begin{figure}[t]
%\centering
%\includegraphics[width=\linewidth]{electrontransport.pdf}
%\caption{\label{fig:electran}Global electron transport flowchart.}
%\end{figure}

The distance to the next hard interaction can once again be sampled with the help of the Woodcock 
raytracing algorithm. For electron transport, however, the total attenuation coefficient, $\mu(E)$, is in fact $
\mu(E,s)$ where the dependence on distance has been made explicit. The dependence on distance is due 
to the fact that electrons will lose energy due to sub-threshold interactions between two successive hard 
interactions. In this application, the approximation is made that the attenuation coefficient is decreasing 
with respect to the particle energy. It is not rigorously true, as shown in~\cite{KAWR00}, but allows for an 
easier treatment since $\mu_{max}(E)$ can be selected with $E$ being the energy at the beginning of the 
electron step.

Energy losses are experienced by the electron during the soft scattering events that are modeled by the 
multiple-scattering step. This energy loss is accounted for with the continuously slowing down 
approximation (CSDA). This approximation states that charged particles are continuously being slowed 
down due to the interactions they go through and that the rate of energy loss is equal to the stopping power 
$S$, given by:
\begin{equation}
S(E) = -\frac{dE}{ds}.
\end{equation}
The implementation of a class II condensed history method only accounts for sub-threshold interactions in the 
CSDA. The use of restricted stopping powers, $L$, is therefore required:
\begin{equation}
L(E,K_{cut}) = \int_0^{K_{cut}}\Sigma(E,E')dE'
\end{equation}
where $\Sigma (E,E')$ is the total macroscopic cross section and $K_{cut}$ is the catastrophic interaction kinetic energy threshold.
Since this release of GPUMCD is $e_s$-centric, the energy loss of a given step is computed with 
\begin{equation}
\Delta E = \int_0^{e_s}L(s)ds .
\end{equation}
The evaluation of the integral can be reduced to
\begin{equation}
\Delta E = L\left(E_0-L(E_0)s/2\right)
\end{equation}
which is EGSnrc's Eq. 4.11.3.

Between two consecutive hard collisions, the electron does not follow a straight line. Bielajew's alternate 
random hinge method~\cite{SALV06} builds on PENELOPE's random hinge method to handle the angular 
deflection and lateral displacement experienced by an electron during an electron step. The random hinge 
method can be described as follows: the electron is first transported by $\zeta e_s$ (where $\zeta$ is again 
a uniformly distributed random number), at which point the electron is rotated by the sampled multiple-
scattering angle described in Sec.~\ref{sec:matcond} and the energy loss is deposited. The electron is then 
transported for the remaining distance equal to $(\zeta -1)e_s$. Bielajew's refinement of the algorithm 
involves randomly sampling the angular deflection either before or after the electron has deposited the 
energy associated with the step.

\subsection{\label{sec:matgpu}GPU implementation}
%si tu veux sauver de l'espace, tu peux supprimer cette partie
In this section, different aspects of the GPU implementation are
detailed. GPUMCD is built with the  CUDA framework from NVIDIA. A general description of the GPU building blocks and  
memory levels can be found in~\cite{HISS09}. Sec.~\ref{sec:matrng} presents the random number generator used in this work
while Sec.~\ref{sec:matmem} describes the memory management.
Finally, Sec.~\ref{sec:matred} addresses the SIMD divergence issue and presents
solutions as well as other optimizations used in GPUMCD.

\subsubsection{\label{sec:matrng}Random number generator}
A pseudo random number generator (PRNG) is already available in the NVIDIA SDK (Software 
Development Kit) for GPU computing. However, it uses a lot of resources which would then be unavailable for the rest of the 
Monte Carlo simulation. For this reason,  a new lightweight PRNG based on the work 
of Marsaglia~\cite{MARS91} was implemented. We use a combined multiply-with-carry (MWC) generator, with recurrence of 
the form
\begin{equation}
x_{n+1}=(a*x_n+c_n) \bmod(b)
\end{equation}
where $a$ is the multiplier and $b$ the base. The carry, $c$, is defined by:
\begin{equation}
c_{n+1}=\lfloor \frac{(a*x_n+c_n)}{b} \rfloor.
\end{equation}
%%tu n
The carry $c$ is naturally computed with integer arithmetic and bases $b$ of $2^{32}$ or $2^{16}$. 
The choice of the multiplier $a$ is not arbitrary: the multiplier is chosen so that $ab-1$ is
safe prime. The period of the 
PRNG with such a multiplier is on the order of $(ab-1)/2$. This PRNG has the advantage of using a small amount of registers. By 
combining two of these generators for each thread, the PRNG uses two 32 bits values for the multipliers and two 32 
bits values for the current state. Ten integer operations (3 $shifts$, 2 $ands$, 2 $mults$ and 3 $adds$) are 
required to generate one new number. It has been shown to pass all tests but one of the TestU01 suite~
\cite{LECU07}, a program designed to test the quality of pseudo random numbers, and to achieve 96.7\% of 
the peak integer bandwidth.

\subsubsection{\label{sec:matmem}Memory management}
GPUMCD is composed of four main arrays: a list of electrons and a list of photons, an array to store 
composition identifiers and another for density values corresponding to a numerical phantom.
During the initialization phase, a call to the \texttt{initParticle} function is made which fills the 
correct particle array (depending on the source particle type) and according to the source type ({\emph{e.g.} with a 
parallel beam source all particles have a direction vector of (0,-1,0)). The particle arrays are allocated only 
once, with size \texttt{MAXSIZE}. The choice of \texttt{MAXSIZE} is graphics card dependent as the particle arrays occupy the better part of the global memory, the largest pool of memory on the graphics card. For instance, on a 1~Gb card, \texttt{MAXSIZE} could be set to $2^{20}$. Only a fraction of the particle array will be filled with source particles, to 
leave room for secondary particle creation. If for instance \texttt{MAXSIZE}/$d$ particles are generated, then 
each primary particle can generate an average of $d$ secondary particle without running out of memory.  The choice of $d$ is therefore important; it is energy dependent and to a lesser extent geometry dependent. For example, with a monoenergetic 15~MeV beam of photons on a 32~cm deep geometry, a value of $d=8$ was found to be sufficient to accommodate all secondary particles. A 
global counter of active particles is kept for both electrons and photons and every time a new particle is 
added the counter is atomically incremented, until it is equal to \texttt{MAXSIZE} after which secondary 
particles are discarded. If the choice of $d$ and \texttt{MAXSIZE} are such that the number of particles generated is lower than \texttt{MAXSIZE}, then no particles are discarded. Atomic operations are a mechanism that ensure that two parallel threads cannot write the same memory location at the same time, therefore discarding one thread's modification. Atomic operations are not ideal on parallel hardware but these two integer values, the current number of particles of both types, should be efficiently cached on GF100 class hardware through the use of the GPU-wide 768~Kb of L2 cache.  After the initialization phase is completed, the 
simulation starts by calling the function to simulate the initial particle type. Subsequently, secondary 
particles of the other type are generated and put inside the particle array for that type. The other type of 
particle will then be simulated, and so on until a relatively low number of particle is left in the stack. This number is user-selectable and could be set to $0$ to simulate every particle. Since a simulation pass always comes with some overhead, it could also be set to a non-zero value to avoid the launching of a simulation pass with few particles to simulate which could have a negligible impact on the final distribution. If this impact is in fact negligible of course depends on the threshold selected.
Photons and electrons are never simulated in parallel but instead the particles of the {\em other} type are placed in their respective array and wait there until the array with the {\em current} type of particle is exhausted. This, in turn, eliminates the divergence due to the photon-electron coupling.
%A flowchart of this design is presented in Fig.~\ref{fig:global}.
Every time a call to a simulation function for electrons or photons is issued, one can consider that every 
particle of that type is simulated and that the counter for that type of particle is reset to 0. Sec.~\ref{sec:matred} 
will show some refinements related to the management of particles.

The other two main arrays, one of composition identifier and one of density values, are stored as 3D textures 
since they are used to represent the volume of voxels. 3D textures reside in memory but they can be efficiently cached for 3D data locality. Every time a particle is moved, the composition and density of the current particle voxel are fetched from the 3D textures. 

Cross section and stopping power data are stored in global memory 
in the form of a 1D texture. The cross section and stopping power data are preinterpolated on a regular grid 
with a value at every 1~keV.

The shared memory usage is relatively limited in this application. For the electrons simulation, some 
specific composition attributes are required (\emph{e.g.} $Z_V$, $Z_G$, etc., in Tab. 2.1 of EGS5) and are 
stored in shared memory. These data values have not been placed in constant memory since that type of memory is best used when all threads of a warp access the same address, which cannot be guaranteed here since these composition attributes are material dependent. Two particles in different materials would therefore request the constant memory at two different addresses, resulting in serialization. No other use for shared memory has been found since the application is by nature stochastic. For instance, we cannot store a portion of the volume since there is no way to know where all the particles of one thread block will be. Similarly, we cannot store a portion of the cross-section data since there is no way to know in which material and with which energy all the particles of one thread block will be.

\subsubsection{\label{sec:matred}Divergence reduction and other optimization strategies}
Every multiprocessor (MP) inside the GPU is in fact a SIMD processor and the SIMD coherency has to be 
kept at the warp level which is the smallest unit of parallelism and is composed of 32 threads. The Monte Carlo simulation being inherently stochastic, it is 
impossible to predict which path a given particle will take and it is therefore impossible to regroup particle 
with the same fate into the same warps. When a warp is divergent, it is split into as many subwarps as there 
are execution paths, leading to a performance penalty. Divergence can be seen when,~\emph{e.g.} two 
particles of the same warp do not require the same number of interactions before exhaustion or do not interact 
in the same way. Some software mechanisms can be employed to reduce the impact of divergence.

The first mechanism consists in performing a stream compaction after $N$ simulation steps, where $N$ is user-defined and corresponds to a number of interactions for photons and to a number of catastrophic events for electrons. For example, if one particle requires 20 
interactions to complete and the others in the warp require less than 5, then some scalar processors (SP) in 
the MP will be idle while they wait for the {\em slow} particle to finish. This first mechanism then artificially limits 
the number of simulation steps a particle can undergo during one pass of the algorithm. After this number of 
steps, every particle that has been completely simulated is removed from the list of particles and the 
simulation is restarted for another $N$ steps, until every particle has been completely simulated. 
The removal of particles is not free and therefore it is not clear if this technique will have a 
positive effect on execution time. The stream compaction is accomplished with the CHAG library~\cite{BILL09}.

A second mechanism, named persistent thread by its creators and {\em pool} in GPUMCD, is taken from the 
world of graphics computing~\cite{AILA09}. In this approach, the minimum number of threads to saturate the 
GPU is launched. These threads then select their workload from a global queue of particles to be simulated. 
Once a thread is done with one particle, it selects the next particle, until all have been simulated. This is 
once again to reduce the impact of the imbalance in the number of interactions per particle.

All the GPU code uses single precision floating point numbers as they offer a significant speed improvement over double precision floating point numbers on graphics hardware. The work of Jia et al.~\cite{JIA10} showed that no floating point arithmetic artifacts were introduced for this type of application.

A notable difference between this work and the work by Jia \emph{et al.} is the way the secondary particles are treated. In the work by Jia, a thread is responsible for its primary particle as well as every secondary particles it creates. This can be a major source of divergence because of the varying number of secondary particles created. On the other hand, GPUMCD does not immediately simulate secondary particles but instead places them in their respective particle arrays. After a given pass of the simulation is over, the arrays are checked for newly created secondary particles and if secondary particles are found, they are simulated. This eliminates the divergence due to the different number of secondary particles per primary particles. Additionally, since  Jia \emph{et al.} use a single thread per primary particles, a thread may be responsible for the simulation of both electrons and photons which can in turn be a major source of divergence. For example, at time $t$, thread $A$ may be simulating a secondary electron while thread $B$ a secondary photon. In other words, they simulate both photons and electrons at the same time. Since both of these particles most likely have completely different code path, heavy serialization will occur. On the other hand, since GPUMCD stores secondary particles in arrays to be simulated later, no such divergence due to the electron-photon coupling occurs.

Finally, GPUMCD can be configured to take advantage of a multi-GPU system. The multi-GPU approach is 
trivial for Monte Carlo simulations: both GPUs execute the \texttt{initParticle} function and simulate their own 
set of particles. After the simulations, the two resulting dose arrays are summed. Linear performance gains 
are expected for simulations when there are enough particles to simulate to overcome the overhead penalty 
resulting from copying input data to two graphics card instead of one.

These optimization mechanisms can be turned on or off at the source code level in GPUMCD.

\subsection{\label{sec:matperf}Performance evaluation}
The performance evaluation of GPUMCD is twofold: a dosimetric evaluation against EGSnrc/DOSXYZnrc 
and an efficiency evaluation against EGSnrc and DPM, the later being a fairer comparison since DPM was 
designed for the same reason GPUMCD was developed, \emph{i.e.} fast MC dose calculations.

All settings for DPM were set to default, notably $K_{cut}$ = 200~keV and $P_{cut}$ = 50~keV. Most 
parameters for DOSXYZnrz (unless otherwise noted) were also set to default, notably \texttt{ESTEPE}=0.25, 
\texttt{$\xi_{max}$}=0.5. In EGSnrc/DOSXYZnrc, the boundary crossing algorithm used was \texttt{PRESTA-I} and 
the electron step algorithm was \texttt{PRESTA-II}, atomic relaxation was turned off as well as bound Compton 
scattering. Values of $P_{cut}$ and $E_{cut}$ were set to 0.01~MeV and 0.7~MeV respectively. The same 
$64^3$ grid with 0.5~$cm^3$ voxels is used for all simulations on all platforms. These values of spatial resolution and number of voxels are insufficient for a clinical calculation with patient specific data; however, they were judged adequate for the benchmarking study conducted here. Preliminary results suggest that 
no loss of efficiency occurs between GPUMCD and other platforms as the number of voxels is increased.

The uncertainty of Monte Carlo simulation results varies as $1/\sqrt{N}$ where $N
$ is the number of histories. 
%For the graphs presented here, enough particles were generated to produce smooth curves where error bars would simply add clutter to the results. For this reason, the graphs are showed without error bars. 
All simulations ran for visualization purposes on GPUMCD were conducted with enough primary particles to achieve a statistical uncertainty of 1\% or less. The same number of particles were then generated in DOSXYZnrc. The graphs are shown without error bars since they would simply add clutter to the results. For execution time comparisons, 4 million particles were generated and simulated for photon 
beams and 1 million particles for electron beams. All sources were modeled as a monoenergetic parallel 
beam. All dose distributions were normalized with respect to $D_{max}$. The efficiency measure, $\epsilon$, 
was used to evaluate the performance of the different implementations:
\begin{equation}
\epsilon = \frac{1}{s^2T},
\end{equation}
where $s$ is the statistical uncertainty and $T$ the computation time. Only voxels with a dose higher than 
$0.2 \cdot D_{max}$ were considered for the statistical uncertainty, yielding the following expression for the uncertainty~\cite{KAWR00X}:
\begin{equation}
s = \frac{1}{N_{0.2}}\sum_{D_{ijk}>0.2D_{max}}\frac{\Delta D_{ijk}}{D_{ijk}},
\end{equation}
where

\begin{equation}
\Delta D^2_{ijk} = \frac{\langle D^2_{ijk}\rangle-\langle D_{ijk}\rangle^2}{N-1},
\end{equation}
for $N$ the number of histories simulated, $\langle D \rangle$ the mean of the random variable $D$ and $N_{0.2}$ the number of voxel with dose higher than $0.2D_{max}$.

For the dosimetric evaluation, a gamma index was used to compare the two dose distributions~\cite{LOW03}. 
The gamma index for one voxel $x$ is defined as
\begin{equation}
\gamma(x) = min\{\Gamma(x,x')\}\forall\{x'\},
\end{equation}
and
\begin{equation}
\Gamma(x,x') = \sqrt{\frac{||x'-x||^2}{\Delta d^2}+\frac{|D(x)-D(x')|^2}{\Delta D^2}},
\end{equation}
where $||x'-x||$ is the distance between voxels $x$ and $x'$, $D(x)$ is the dose value of voxel $x$, $\Delta 
d$ is the distance tolerance value and $\Delta D$ is the dose tolerance value. The criteria for acceptable 
calculation is defined for a gamma index value below or equal to 1.0.

All simulations were run on the same PC comprising an Intel Core i7 860 and a NVIDIA GeForce GTX480 
graphics card. DPM and DOSXYZnrc do not natively support multi-core architectures and have not been 
modified. GPUMCD, unless running in multi-GPU configuration, uses only one processor core.

\section{\label{sec:results}Results}
Several slab-geometry phantoms are described in Sec.~\ref{sec:dosres} . In Sec.~\ref{sec:speedres}, the 
execution times and gains in efficiency are reported. A multi-GPU implementation is also tested and corresponding 
acceleration factors are presented.
\subsection{\label{sec:dosres}Dosimetric results}
For slab geometries, central axis percent depth dose (PDD) curves as well as dose profiles at different 
depths or isodose curves are presented. Gamma indices are calculated in the entire volume of voxels and 
the maximum and average values are reported. The number of voxels with a gamma value higher than 1.0 
and 1.2 are also detailed in order to evaluate discrepancies in dose calculation. The gamma criteria used is 
set to 2\% and 2~mm which is a generally accepted criteria for clinical dose calculation~\cite{DYKE99}. In 
every gamma index summary table: 1) $\gamma^{max}$ is the maximum gamma value for the entire 
volume of voxel; 2) $\gamma_{c}^{avg}$ is the average gamma value for voxels with $D>cD_{max}$; 3) $
\Sigma D_{c}$ is the number of voxels with $D>cD_{max}$ ; 4)  $\Sigma \gamma_{c}>g$ is the number (proportion) of 
voxels with $D>cD_{max}$ and $\gamma>g$.
%% number of voxels or proportion of voxels??

The first slab geometry is a simple water phantom. GPUMCD treats homogeneous phantoms as heterogeneous phantoms; the 
particles are transported as if evolving inside a heterogeneous environment and therefore  there is no 
gain in efficiency resulting from homogeneous media. For this geometry, the results for an 
electron beam and a photon beam are presented in Fig.~\ref{fig:15mevwater} and gamma results in Tab.~\ref{tab:resgamma15mevwater}.
%pas certain commentaire hugo ici

%water
\begin{figure*}[htp]
\centering
\includegraphics[width=\linewidth]{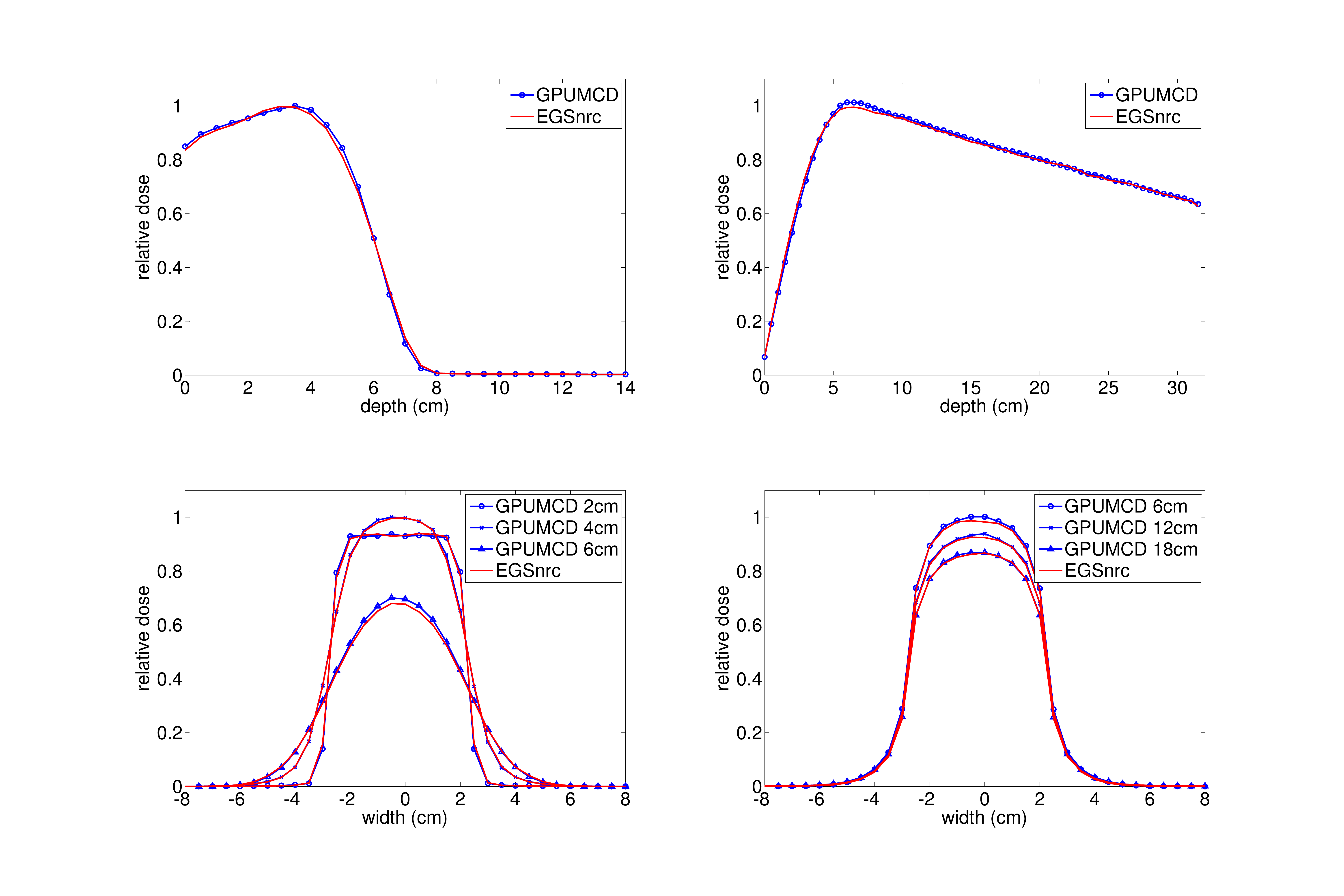}
\caption{\label{fig:15mevwater} PDD (top) and profile (bottom) of 15~MeV electron (left) and photon (right) 
beams on water.}
\end{figure*}

\begin{table}[htp]
\begin{center}
\begin{tabular}{lccccc}
\hline
Particles & $\gamma^{max}  $  & $\gamma_{0.2}^{avg}$  &  $\Sigma D_{0.2}$  &  $\Sigma \gamma_{0.2} 
>1.0 $ & $\Sigma \gamma_{0.2}>1.2$ \\
\hline  \hline
Electrons & 1.24 &0.29 & 1750& 37~(2\%) & 2~($\sim$0\%)\\
Photons & 1.27 &0.18 & 7500& 143~(2\%) & 19~($\sim$0\%)\\
\hline
\end{tabular}
\caption{\label{tab:resgamma15mevwater}Gamma criteria summary for a 15~MeV  beam on water.}
\end{center}
\end{table}

%The second geometry is composed of water and lung layers. For electrons, the lung layer is 4~cm thick 
%and is placed between 3~cm and 7~cm; for photons the lung layer is 5~cm thick and is placed between 
%10~cm and 15~cm. The PDD and profile curves for the electron and photon beams are presented in Fig.~
%\ref{fig:15mevlung} while the and gamma results are presented in Tab.~\ref{tab:resgamma15mevlung}.
%
%%lung
%
%\begin{figure*}[htp]
%\centering
%\includegraphics[width=\linewidth]{combined15mevlung.pdf}
%\caption{\label{fig:15mevlung}PDD (top) and profile (bottom) of a (left) 15~MeV electron beam on water 
%with 4~cm lung layer between 3~cm and 7~cm and (right) 15~MeV photon beam on water with 5~cm lung 
%layer between 10~cm and 15~cm.}
%\end{figure*}
%
%\begin{table}[htp]
%\begin{center}
%\begin{tabular}{lccccc}
%\hline
%Particle & $\gamma^{max}  $  & $\gamma_{0.5}^{avg}$  &  $\Sigma D_{0.5}$  &  $\Sigma \gamma_{0.5} 
%>1.0 $ & $\Sigma \gamma_{0.5}>1.2$ \\
%\hline  \hline
%Electron & 1.26 &0.34 & 1185& 1~($\sim$0\%) & 0~(0\%)\\
%Photon & 1.60 &0.20 & 5413& 97~(2\%) & 36~($\sim$0\%)\\
%\hline
%\end{tabular}
%\caption{\label{tab:resgamma15mevlung}Gamma criteria summary for a 15~MeV  beam on water with a 
%lung layer.}
%\end{center}
%\end{table}

The second geometry is composed of a lung box inside a volume of water. It is designed to demonstrate the performance of 
GPUMCD with lateral and longitudinal disequilibrium conditions. For electrons, the lung box is 4~cm long, 
4.5~cm wide and starts at a depth of 3~cm; for photons the lung box is 6.5~cm long, 
4.5~cm wide and starts at a depth of 5.5~cm. In both cases, the box is centered on the central axis. For this geometry, PDD and 
isodose plots  are presented in Fig.~\ref{fig:15mevlungbox} for the electron and photon beams while the 
gamma results are presented in Tab.~\ref{tab:resgamma15mevlungbox}.

%lung

\begin{figure*}[htp]
\centering
\includegraphics[width=\linewidth]{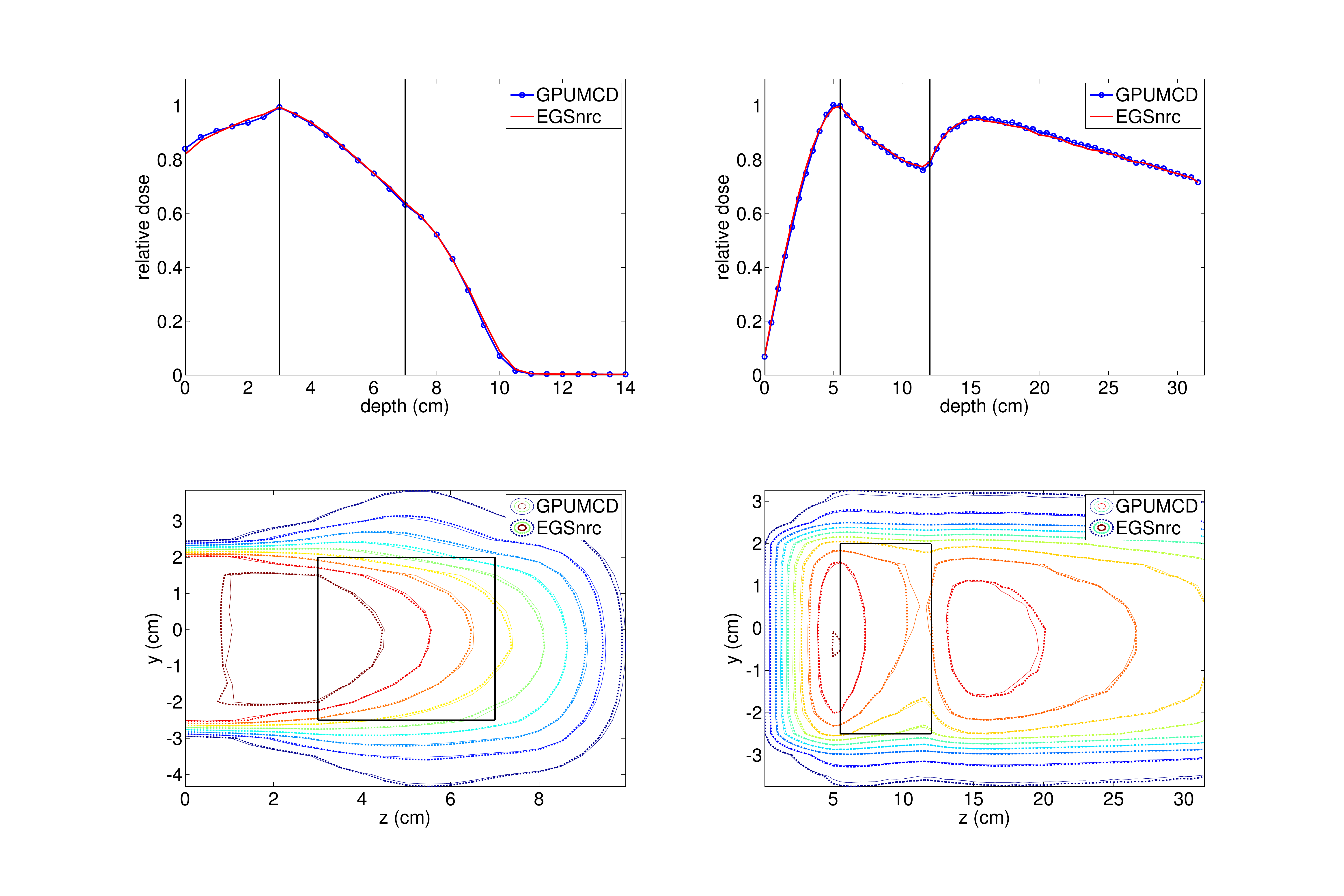}
\caption{\label{fig:15mevlungbox}PDD (top) and isodose (bottom) of a (left) 15~MeV electron beam on 
water with a 4~cm long, 4.5~cm wide box of lung at a depth of 3~cm and (right) 15~MeV photon beam 
on water with a 6.5~cm long, 4.5~cm wide box of lung at a depth of 5.5~cm.}
\end{figure*}

\begin{table}[htp]
\begin{center}
\begin{tabular}{lccccc}
\hline
Particles & $\gamma^{max}  $  & $\gamma_{0.2}^{avg}$  &  $\Sigma D_{0.2}$  &  $\Sigma \gamma_{0.2} 
>1.0 $ & $\Sigma \gamma_{0.2}>1.2$ \\
\hline  \hline
Electrons & 1.40 &0.32 & 2366& 46~(2\%) & 2~(0\%)\\
Photons & 1.30 &0.19 & 7522& 122~(2\%) & 19~($\sim$0\%)\\
\hline
\end{tabular}
\caption{\label{tab:resgamma15mevlungbox}Gamma criteria summary for a 15~MeV  beam on water with a 
lung box.}
\end{center}
\end{table}

%The third geometry is composed of water, aluminum and lung. The slabs, for electrons, are arranged as 
%such: 2~cm of water, 1~cm of aluminum, 3~cm of lung followed by water. For photons the geometry is: 3~cm 
%of water, 2~cm of aluminum, 7~cm of lung followed by water. The result for the electron beam and photon 
%beam are presented in Fig.~\ref{fig:15meviccr} and gamma results in Tab.\ref{tab:resgamma15meviccr}.
%
%% iccr
%\begin{figure*}[htp]
%\centering
%\includegraphics[width=\linewidth]{combined15meviccr.pdf}
%\caption{\label{fig:15meviccr}PDD (top) and profile (bottom) of a (left)  15~MeV electron beam on a 
%phantom composed of  2~cm of water, 1~cm of aluminum, 3~cm of lung followed by water and (right) 
%15~MeV photon beam on a phantom composed of 3~cm of water, 2~cm of aluminum, 7~cm of lung followed 
%by water.}
%\end{figure*}
%
%
%\begin{table}[htp]
%\begin{center}
%\begin{tabular}{lccccc}
%\hline
%Particle & $\gamma^{max}  $  & $\gamma_{0.5}^{avg}$  &  $\Sigma D_{0.5}$  &  $\Sigma \gamma_{0.5} 
%>1.0 $ & $\Sigma \gamma_{0.5}>1.2$ \\
%\hline  \hline
%Electron & 1.82 &0.58 & 872& 106~(12\%) & 43~(5\%)\\
%Photon & 2.09 &0.24 & 3956& 187~(5\%) & 95~(3\%)\\
%\hline
%\end{tabular}
%\caption{\label{tab:resgamma15meviccr}Gamma criteria summary for a 15~MeV  beam on a water 
%phantom with aluminum and lung layers.}
%\end{center}
%\end{table}

The third geometry is composed of soft tissue, bone and lung. The slabs, for electrons, are arranged as such: 
2~cm of soft tissue, 2~cm of lung, 2~cm of bone followed by soft tissue. For photons the geometry is: 3~cm of soft tissue, 
2~cm of lung, 7~cm of bone followed by soft tissue. The results for the electron and photon beams are 
presented in Fig.~\ref{fig:15mevbone} and gamma results in Tab.~\ref{tab:resgamma15mevbone}.

% iccr
\begin{figure*}[htp]
\centering
\includegraphics[width=\linewidth]{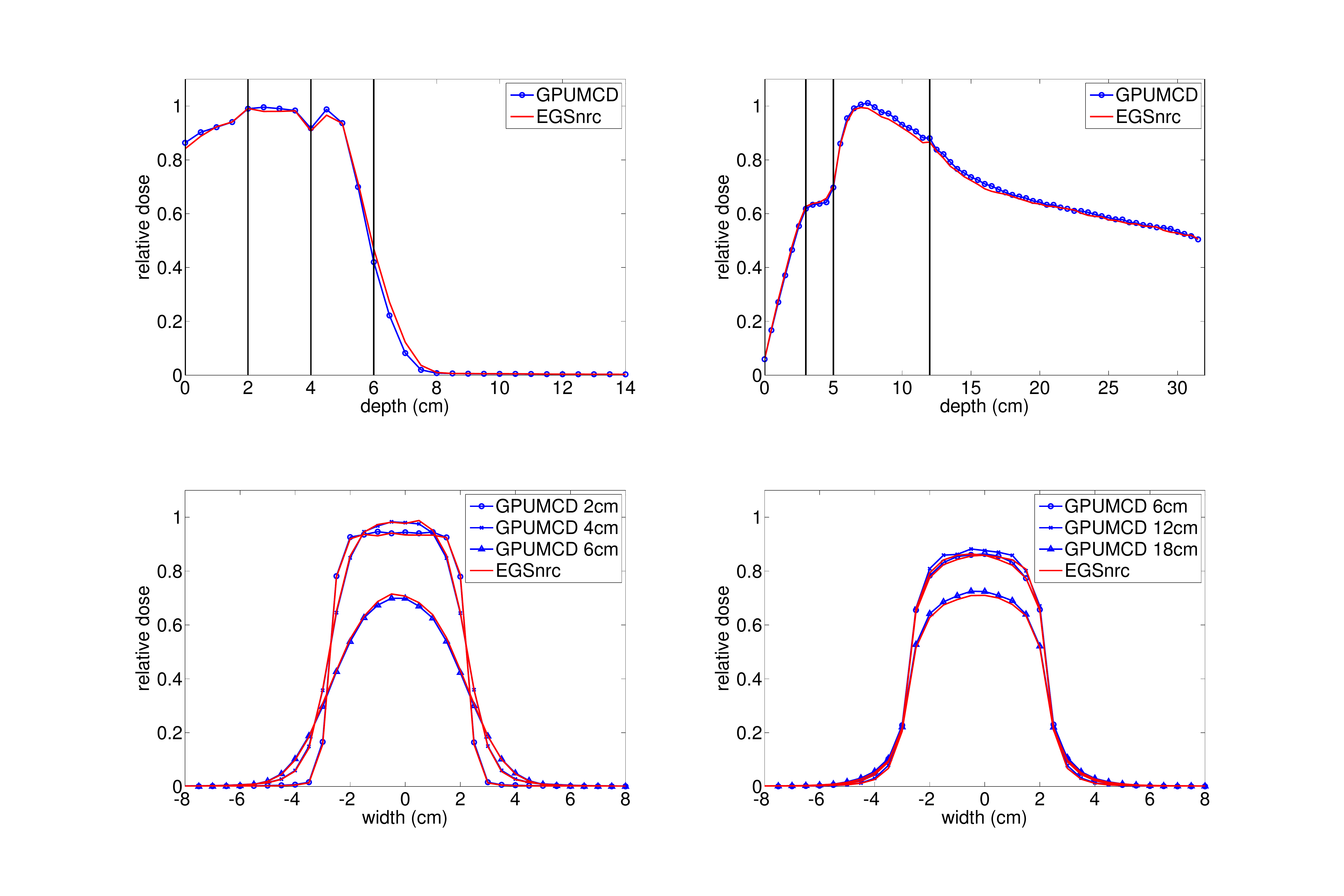}
\caption{\label{fig:15mevbone}PDD (top) and profile (bottom) of a (left)  15~MeV electron beam on a 
phantom composed of  2~cm of soft tissue, 2~cm of bone, 2~cm of lung followed by soft tissue and (right) 15~MeV 
photon beam on a phantom composed of 3~cm of soft tissue, 2~cm of lung, 7~cm of bone followed by soft tissue. 
}
\end{figure*}

\begin{table}[htp]
\begin{center}
\begin{tabular}{lccccc}
\hline
Particles & $\gamma^{max}  $  & $\gamma_{0.2}^{avg}$  &  $\Sigma D_{0.2}$  &  $\Sigma \gamma_{0.2} 
>1.0 $ & $\Sigma \gamma_{0.2}>1.2$ \\
\hline  \hline
Electrons & 1.84 &0.34 & 1700& 138~(8\%) & 54~(3\%)\\
Photons & 1.31 &0.22 & 7220& 12~($\sim$0\%) & 0~(0\%)\\
\hline
\end{tabular}
\caption{\label{tab:resgamma15mevbone}Gamma criteria summary for a 15~MeV  beam on a  
phantom with layers of soft tissue, bone and lung.}
\end{center}
\end{table}

\subsection{\label{sec:speedres}Execution time and efficiency gain}
In this section, the absolute execution times as well as the overall speed and efficiency of GPUMCD is evaluated and compared to EGSnrc and 
DPM. In the following tables, $Geom_1$ is the water phantom, $Geom_2$ is the water-lung phantom and 
$Geom_3$ is the tissue-lung-bone phantom. $T_{EGSnrc}$ and $T_{DPM}$ are respectively the EGSnrc and 
DPM execution times. For the efficiency measurement, $\epsilon$, the fastest execution time of GPUMCD 
(with or without the optimization presented in Sec.~\ref{sec:matred}) has been used. The acceleration factor 
and efficiency improvement, namely $A$ and $A_{\epsilon}$, also use the fastest time for GPUMCD. The 
acceleration factor is defined as
\begin{equation}
A = \frac{T_{ref}}{T_{GPUMCD}},
\end{equation}
and the efficiency improvement as
\begin{equation}
A_{\epsilon} = \frac{\epsilon_{GPUMCD}}{\epsilon_{ref}}.
\end{equation}

Tab.~\ref{tab:resabsexec} presents the absolute execution times for all simulations with both electron and photon beams.

\begin{table}[htp]
\begin{center}
\begin{tabular}{c|ccc}
\hline\hline
Particles & $Geom_1$ & $ Geom_2 $ & $Geom_3$ \\
&\multicolumn{3}{c}{(seconds)}\\
\hline \hline
Electrons & 0.118 & 0.147 & 0.162  \\
Photons & 0.269 & 0.275 & 0.366 \\
\hline
\end{tabular}
\caption{\label{tab:resabsexec} Absolute execution times, in seconds, for GPUMCD in the three geometries for 1M electrons and 4M photons using 15~MeV monoenergetic beams. }
\end{center}
\end{table}

Tab.~\ref{tab:resexec} presents the acceleration results for photon and electron beams where 4M photons and 1M electrons are simulated. 

%\begin{table}[htp]
%\begin{center}
%\begin{tabular}{cc|cc|cc}
%\hline\hline
%&&\multicolumn{2}{c|}{Electron} & \multicolumn{2}{|c}{Photon}\\
%\hline
% & &  DPM  & EGSnrc & DPM & EGSnrc\\
%\hline \hline
%%&$T$ (s) & 1.80 & 157.3 & 403 \\
%\multirow{2}{*}{$Geom_1$}&$A$  & 87.1 & 223 & 40.0 & 234\\
%&$A_\epsilon$ & 203 & 372 & 108 & 376\\
%\hline
%%&$T$ (s) & 1.75 & 153.8 & 375 \\
%\multirow{2}{*}{$Geom_2$}&$A$  & 87.8 & 214 & 38.7 & 204\\
%&$A_\epsilon$  & 180 & 328 & 104 & 374\\
%\hline
%%&$T$ (s) & 2.27 & 96.2 & 448 \\
%\multirow{2}{*}{$Geom_3$}&$A$  & 42.4 & 197 & 41.9 & 228\\
%&$A_\epsilon$  &85.7 & 324 & 108 & 376\\
%\end{tabular}
%\caption{\label{tab:resexec} Accelerations for 1M 15~MeV electrons and 4M 15~MeV photons on the 
%three geometries presented. Absolute execution times of 1.80~s, 1.75~s and 2.27~s are found with 
%GPUMCD for $Geom_1$, $Geom_2$ and $Geom_3$ respectively for a photon beam and 0.94~s, 1.05~s 
%and 1.14~s for an electron beam. }
%\end{center}
%\end{table}

\begin{table}[htp]
\begin{center}
\begin{tabular}{cc|cc|cc}
\hline\hline
&&\multicolumn{2}{c|}{Photons} & \multicolumn{2}{|c}{Electrons}\\
\hline
 & &  DPM  & EGSnrc & DPM & EGSnrc\\
\hline \hline
%&$T$ (s) & 1.80 & 157.3 & 403 \\
\multirow{2}{*}{$Geom_1$}&$A$  & 453 & 1161 & 246 & 1510\\
&$A_\epsilon$ & 515 & 1174 & 291 & 1740\\
\hline
%&$T$ (s) & 1.75 & 153.8 & 375 \\
\multirow{2}{*}{$Geom_2$}&$A$  & 433 & 1058 & 210 & 1264\\
&$A_\epsilon$  & 474 & 1305 & 237 & 1403\\
\hline
%&$T$ (s) & 2.27 & 96.2 & 448 \\
\multirow{2}{*}{$Geom_3$}&$A$  & 203 & 947 & 225 & 1231\\
&$A_\epsilon$  &220 & 1050 & 242 & 1359\\
\end{tabular}
\caption{\label{tab:resexec} Acceleration factors for 1M 15~MeV electrons and 4M 15~MeV photons on the three 
geometries presented. Absolute execution times of 0.269~s, 0.275~s and 0.366~s were found with GPUMCD 
for $Geom_1$, $Geom_2$ and $Geom_3$ respectively for a photon beam and 0.118~s, 0.147~s and 
0.162~s for an electron beam. }
\end{center}
\end{table}
Tab.~\ref{tab:resdivred} presents the results of the divergence reduction methods and acceleration strategies 
presented in Sec.~\ref{sec:matred} for electron and photon beams. In the following table 
$A_{comp}$, $A_{pool}$ represent the acceleration factors with respect to the base 
configuration execution time ($T_{base}$), with the stream compaction or {\em pool} divergence reduction methods enabled, respectively.

\begin{table}[htp]
\begin{center}
\begin{tabular}{cccc}
\hline\hline
Particles & $T_{base}$ & $ A_{comp} $ & $A_{pool}$ \\
\hline \hline
Electrons & 0.185 & 1.06 & 1.26  \\
Photons & 0.275 & 0.85 & 0.98 \\
\hline
\end{tabular}
\caption{\label{tab:resdivred} Acceleration results of the different strategies presented in Sec.~\ref{sec:matred} for the second geometry ($geom_2$).}
\end{center}
\end{table}

As detailed in Sec.~\ref{sec:matred}, GPUMCD can take advantage of a system with multiple GPUs. The 
impact of parallelizing GPUMCD on two GPUs is presented in Tab.~\ref{tab:resmulti} where $TPH$ (time per history) is the execution time normalized to one complete history including a primary particle and all its secondary particles, $TPH_{s}$ and 
$TPH_{d}$ are the execution time in single and dual GPU mode, respectively. The 
graphics cards used for this test are a pair of Tesla C1060. The reported execution times include every memory transfer to and from the graphics cards.

\begin{table}[htp]
\begin{center}
\begin{tabular}{cccc}
\hline\hline
& $TPH_{s}$ & $ TPH_{d} $ & $A_{d/s}$\\
& ($\mu$s) &  ($\mu$s)  & \\
\hline \hline
Electrons & 5.43 & 2.82 & 1.92\\
Photons & 5.20 & 2.74 & 1.90\\
\hline
\end{tabular}
\caption{\label{tab:resmulti} Execution times and acceleration factors achieved with a multi-GPU 
configuration.}
\end{center}
\end{table}

\section{\label{sec:discussion}Discussion}
\subsection{\label{sec:discdosi} Dosimetric evaluation}
Figures~\ref{fig:15mevwater}~-~\ref{fig:15mevbone} and Tables~\ref{tab:resgamma15mevwater}~-~\ref{tab:resgamma15mevbone} present the dosimetric evaluation performed between GPUMCD and 
EGSnrc.
An excellent agreement is observed for homogeneous phantoms with both
photon and electron beams, except for the end of the build-up region
where small differences of less than 2\% can be observed. 
In both cases, 
the rest of the curve as well as the overall range of the particles are not affected.

In heterogeneous simulations with low density materials, the agreement is also excellent. Differences of up to 
2\% can be seen in the build-up region for an incoming electron beam. These differences again do not affect 
the range of the particles. For a photon beam impinging on the water-lung phantom, the dose is in 
good agreement within as well as near the heterogeneity.

In heterogeneous simulations with a high-Z material, differences of up to 2\% are found with an impinging 
electron beam and no differences above 1\% are found for the photon beam in the presented PDD while 
differences of up to 1.5\% are found in the profiles. 

The overall quality of the dose comparison can be evaluated with the gamma value results presented in 
Tab.~\ref{tab:resgamma15mevwater}~to~\ref{tab:resgamma15mevbone}. Results suggest that the code is 
suitable for clinical applications as the dose accuracy is within the 2\%/2~mm criteria, for all cases but one, which is below the 
recommended calculation accuracy proposed by Van Dyk~\cite{DYKE99}. For electron beams, the gamma 
criteria has been well respected in the first two geometries where at most 2\% of the significant voxels failed. 
More pronounced differences have been found in the last geometry including a high-Z material where 8\% 
of the significant voxels fail the gamma test. However, only 3\% of all significant voxels fail that test with a 
value higher than 1.2, indicating that the critera are slightly too strict. Indeed, with a 2.5\%-2.5mm, 
3\% of all significant voxels fail the test. For photon beams, all geometries meet the gamma criteria for 98
\% or more of all significant voxels.

\subsection{\label{sec:disctime} Execution times}
The differences in calculation efficiency between GPUMCD and the other two codes are due to three 
distinctive reasons: 1) differences in hardware computational power,
2) differences in the programming model and its adaptation to hardware and 3) differences in the physics simulated and approximations 
made. As an operation-wise equivalent CPU implementation of GPUMCD does not exist, these 
factors cannot be evaluated individually.

Also, the simulation setup for the test cases uses monoenergetic beams, which can favor GPU implementations. A naive GPU implementation of polyenergetic beams would lead to additional stream divergence. However, simple measures such as grouping particles with similar energy values would reduce this divergence. Future work will explore this issue in the context of clinical calculations done with polyenergetic beams. 

Tab.~\ref{tab:resabsexec} shows the absolute execution times for electron and photon beams with GPUMCD. Execution times of less than 0.2~s are found for all electron cases. The execution time is higher in the simulation with lung when compared to the simulation in water likely due to the fact that more interactions are necessary. The execution time is higher in the simulation with bone because a larger number of fictitious interactions will be encountered. Similar conclusions can be taken in the cases with photon beams in which the first two simulations require less than 0.3~s and the simulation with a bone slab required more than 0.36~s. 

Tab.~\ref{tab:resexec} shows the acceleration results comparing GPUMCD to DPM and EGSnrc. It can be 
seen that GPUMCD is consistently faster than DPM, by a factor of at least 203x for photon beams and 210x 
for electron beams. The disadvantage of using the Woodcock raytracing algorithm is apparent in Tab~\ref{tab:resexec} where the geometry featuring a heavier than water slab has the lowest acceleration factor. 
An acceleration of more 1200x is observed when comparing to EGSnrc for both electron beans and more than 940x for photon beams.

DPM and EGSnrc both employ much more sophisticated electron step algorithms compared to GPUMCD. 
Improvements to the electron-step algorithm in GPUMCD will most likely yield greater accelerations, as 
photon transport was shown to be much faster than the reported coupled transport results~\cite{HISS10b}, 
as well as improved dosimetric results.

Tab.~\ref{tab:resdivred} presents the execution times and acceleration factors for the different divergence 
reduction methods and optimizations presented in Sec.~\ref{sec:matred}. Gains of up to 26\% for electron 
beams are observed. However, acceleration factors below 1 are found for photon beams, showing that electron beams 
are more divergent. These results show that the GPU architecture is not ideal for Monte Carlo simulations but that the hardware is able to cope well with this divergence. The mechanisms employed in GPUMCD to reduce this penalty have either had a negative impact on the execution time or a small positive impact that is perhaps not worth the loss in code readability and maintainability.

Tab.~\ref{tab:resmulti} details the gain that can be obtained by executing GPUMCD on multiple graphics 
card. GPUMCD presents a linear growth with respect to the number of GPUs. This is expected as there is a 
relatively small amount of data transferred to the GPUs compared
to the amount of computations required
 in a typical simulation. As the number of histories is reduced, so is
 the acceleration factor.

Efficiency improvements can be achieved with variance reduction techniques (VRT). Several 
methods have been published to improve the calculation efficiency of
EGSnrc for instance~\cite{KAWR00,WULF08}. A fair comparison 
of GPUMCD with other Monte Carlo packages would ideally involve an optimal usage 
of the codes. In this paper, no VRT was used for the comparisons. It is not clear yet how VRT could potentially 
be implemented in GPUMCD. In the eventuality that VRT are equally applicable to each code architecture, the 
efficiency improvements reported in this paper would remain roughly
the same. Otherwise, GPUMCD might suffer from a performance penalty
compared to VRT-enabled codes.

\section{Conclusion and future work}
In this paper, a new fully coupled GPU-oriented Monte Carlo dose calculation 
platform was introduced. The accuracy of the code was evaluated with various geometries and compared to EGSnrc, a thoroughly validated Monte Carlo package. The overall speed of the platform was compared to DPM, an established fast Monte Carlo simulation package for dose calculations. 
For a 2\%-2mm gamma criteria, the dose comparison is in agreement for 98\% or more of all significant voxels, except in one case:  
the tissue-bone-lung phantom with an electron beam where 92\% of significant voxels pass the gamma 
criteria. These results suggest that GPUMCD is suitable for clinical use.

The execution speed achieved by GPUMCD, at least two orders of magnitude faster than DPM, let envision the use 
of accurate of Monte Carlo dose calculations for numerically intense applications such as 
 IMRT  or arc therapy optimizations. A 15~MeV electron beam dose calculation in water can be performed 
in less than 0.12~s for 1M histories while a photon beam calculation takes less than 0.27~s for 4M histories.

GPUMCD is currently under active development. Future work will include the ability to work with phase space and 
spectrum files as well as the integration and validation of CT-based phantoms for dose calculation.
%Alternatives to the current electron step length selection will also be investigated, which should further 
%improve the execution speed.
Research towards variance reduction techniques that are compatible with the GPU architecture is also planned.

\section*{Acknowledgment}
This work was supported by Natural Sciences and Engineering Research Council of Canada (NSERC). 
Dual GPU computations were performed on computers of the Réseau québécois de calcul de haute 
performance (RQCHP). Multiple scattering data has been 
provided by the National Research Council of Canada (NRC).

\section*{References}
\bibliographystyle{myunsrt}
\bibliography{biblio}% Produces the bibliography via BibTeX.

%L493: ? check the name of the first author C'est l'article de CHarlie Ma http://iopscience.iop.org/0031-9155/45/9/303 et c'est bien comme ca que c'est écrit.
% L565: check the reference format ? Je sais pas cqui veut.
%L570: page number? ? trouver la page dans le bouquin.

\end{document}